\documentclass[aps,onecolumn,nofootinbib,groupedaddress,floatfix]{revtex4-1}

\usepackage{placeins}
\usepackage{booktabs}
\usepackage{dcolumn}
\usepackage{array}
\usepackage{longtable}
\usepackage{float}
\usepackage{hyphenat}
\usepackage{natbib}
\usepackage{hyperref}

\usepackage{amsmath,amsfonts,amsthm,amssymb}
\usepackage{bm,graphicx,graphics,color}
\usepackage{epsf,epsfig}
\usepackage[all]{xy}
\newcommand*{\B}[1]{\ifmmode\bm{#1}\else\textbf{#1}\fi}
\begin{document}

\title{Cosmological Solutions for the Geometrical Scalar-Tensor with the Potential Determined by the Noether Symmetry Approach}

\author{Adriano B. Barreto}
\email{adriano.barreto@caxias.ifrs.edu.br}
\affiliation{Instituto Federal de Educa\c{c}\~{a}o, Ci\^{e}ncia e Tecnologia do Rio Grande do Sul, \textit{Campus} Caxias do Sul, Brazil}
\affiliation{Departamento de F\'{i}sica, Universidade Federal do Paran\'{a}, Curitiba, Brazil}

\author{Gilberto M. Kremer}
\email{kremer@fisica.ufpr.br }
\affiliation{Departamento de F\'{i}sica, Universidade Federal do Paran\'{a}, Curitiba, Brazil}

\begin{abstract}
In this work we consider a scale-tensor theory in which the space-time is endowed with a Weyl integrable geometrical structure due to the Palatini variational method. Since the scalar field has a geometrical nature (related to non-metricity), the theory is known as \textit{Geometrical Scalar-Tensor}. On the framework of Weyl transformations, a non-minimally coupled scalar-tensor theory on the Jordan frame corresponds to a minimally coupled Einstein-Hilbert action on the Einstein frame. The scalar potential is selected by the Noether symmetry approach in order to obtain conserved quantities for the FRW cosmological model. Exact solutions are obtained and analyzed in the context of the cosmological scenarios consistent with an expanding universe. A particular case is matched in each frame and the role of scalar field as a dark energy component is discussed.
\end{abstract}
\keywords{Noether symmetry; exact solutions; FRW spacetime; scalar field cosmologies.}
\maketitle 
\section{Introduction}
Although Einstein’s theory of gravity is constantly being supported by current observational data \cite{Will2014}, recent issues such as the accelerated expansion of the Universe and the possible existence of dark matter, can not be fully explained only based on general relativity. In this sense, there have been considerable efforts in the development of alternative theories to Einstein's theory \cite{Yunes2013}. In particular, there is a great interest in investigating new possibilities that include changes in the theory of general relativity \cite{Capozziello2011, Capozziello2014}. 

Scalar tensor (ST) theories are among the proposed extensions of Einstein's theory \cite{Brans:1961sx, DKSen-ScalarTensor, KADunn-ScalarTensor, Faraoni:2019sxw}. A geometrical approach to theories with non-minimal coupling is particularly interesting. According to it, by considering the Palatini variational method a not necessarily Riemannian compatibility condition between the metric tensor and the affine connection -- initially taken as independent variables -- is obtained \cite{Burton1997, Kozak:2018vlp}. Furthermore, it was shown that the geometry that naturally appears when a symmetric affine connection is regarded is the so called Integrable Weyl Geometry, where the scalar field takes part together with the metric tensor in the description of the gravitational field. This brings a geometrical origin for the scalar field present in the theory, which naturally define the known as \textit{geometrical scalar-tensor} theories \cite{Almeida:2013dba, Pucheu:2016act}.

The Noether symmetry approach is regarded in order to select a potential term whose the model can present a conserved quantity \cite{Capozziello:1998nd, Paliathanasis:2012at, Camci:2018apx}. Such conserved quantity will imply in the existence of a cyclic variable useful to found exact solutions for the field equation. This approach has already been used in choosing for models of tensor-scalar theories that could describe the dynamics of the expanding universe, as dark energy models \cite{deSouza:2008nj,deSouza:2010ym,deSouza:2011tx,deSouza:2013uu}.

Therefore, this work is organized as follows: in the Section \ref{secModel} the action of the model is presented. The field equations from a point-like Lagrangian for a flat Friedmann-Robertson-Walker (FRW) metric are derived in the section \ref{secPointlike}. Noether symmetry approach is considered in the section \ref{secNoether} to specify the self-interacting potential of the scalar field. In the section \ref{secSolution} the field equations are integrated via constant of motion and cyclic variable and the comparison between the cosmological solutions in each frame are fulfilled in the section \ref{secCosmologicalSolutions}. The particular case $\omega = 1/2$ is seen in section \ref{secParticularCase}. The paper is closed with the conclusions in the section \ref{secConclusions}.

In this work we adopt the natural units $8\pi G = \hbar = c = 1$ and the metric signature $(+, -, -, -)$.
\section{The Model and The Weyl Transformations\label{secModel}}
Let us  start by writing  the action on the \textit{Einstein frame} as
\begin{equation}
    \mathcal{S}_{EF} = \int d^{4}x \sqrt{-\bar{g}}\left[ \bar{R} + \omega \phi^{,\alpha}\phi_{,\alpha} - V\left(\phi\right)\right] \label{Sef}
\end{equation}
where $\phi$ represents a scalar field, $V(\phi)$ its self-interacting potential and $\omega$ a dimensionless coupling constant. Furthermore, $\bar{R} \doteq \bar{g}^{\mu\nu}\bar{R}_{\mu\nu}$ is the scalar curvature (or the Ricci scalar), calculated with the affine connection $\Gamma^{\alpha}_{\mu\nu}$, which is given by
\begin{equation}
\Gamma^{\alpha}_{\mu\nu} = \lbrace^{\alpha}_{\mu\nu}\rbrace - \frac{1}{2}g^{\alpha\beta}\left( g_{\mu\beta}\partial_{\nu}\phi +  g_{\nu\beta}\partial_{\mu}\phi -  g_{\mu\nu}\partial_{\beta}\phi\right). \label{WeylConnection}
\end{equation}
Here $\lbrace^{\alpha}_{\mu\nu}\rbrace = \frac{1}{2}g^{\alpha\beta}\left( \partial_{\nu}g_{\mu\beta}+  \partial_{\mu}g_{\nu\beta} -  \partial_{\beta}g_{\mu\nu}\right)$ represent the Christoffel symbols (Levi-Civita connection). We can apply the Weyl transformation to carry this action on the Jordan frame as follows:
\begin{eqnarray}
\bar{g}_{\mu\nu} & = & e^{-\phi}g_{\mu\nu} \nonumber\\
\bar{g}^{\mu\nu} & = & e^{\phi}g^{\mu\nu} \nonumber\\
\sqrt{-\bar{g}} & = & e^{-2\phi}\sqrt{-g}, \nonumber
\end{eqnarray}
so that eq.(\ref{Sef})  becomes
\begin{eqnarray}
\mathcal{S}& = & \int d^{4}x \sqrt{-\bar{g}}\left[ \bar{g}^{\mu\nu}\bar{R}_{\mu\nu} + \omega \bar{g}^{\mu\nu}\partial_{\mu}\phi\partial_{\nu}\phi - V\left(\phi\right)\right] \nonumber \\
\therefore \mathcal{S}& = & \int d^{4}x \sqrt{-g}e^{-2 \phi}\left[ e^{\phi}g^{\mu\nu}R_{\mu\nu} + e^{\phi}\omega g^{\mu\nu}\partial_{\mu}\phi\partial_{\nu}\phi - V\left(\phi\right)\right] \nonumber \\
\therefore \mathcal{S}& = & \int d^{4}x \sqrt{-g}e^{-\phi}\left[ g^{\mu\nu}R_{\mu\nu} + \omega g^{\mu\nu}\partial_{\mu}\phi\partial_{\nu}\phi - e^{-\phi}V\left(\phi\right)\right] \nonumber.
\end{eqnarray}
Note that  $\bar{R}_{\mu\nu} = R_{\mu\nu}$, because the affine connection $\Gamma^{\alpha}_{\mu\nu}$ is invariant on Weyl transformation.  Therefore, we have obtained the action on the \textit{Jordan frame} as
\begin{equation}
\mathcal{S}_{JF} = \int d^{4}x \sqrt{-g}e^{-\phi}\left[ R + \omega \phi^{,\alpha}\phi_{,\alpha} - e^{-\phi}V\left(\phi \right)\right]. \label{Sjf}
\end{equation}

By performing the Palatini variation with respect to the metric tensor $g_{\mu\nu}$, we obtain the following field equations \cite{Almeida:2013dba},
\begin{equation}
R_{\mu\nu} - \frac{1}{2}R g_{\mu\nu} = - T^{(\phi)}_{\mu\nu} \label{EinsteinFieldEqs}
\end{equation}
where $T^{(\phi)}_{\mu\nu}$ is the energy-momentum tensor of the scalar field
\begin{equation}
    T^{(\phi)}_{\mu\nu} =  \omega \left(\partial_{\mu}\phi \partial_{\nu}\phi -\frac{1}{2}\phi^{,\alpha}\phi_{,\alpha}g_{\mu\nu}\right) + \frac{1}{2}e^{-\phi}V g_{\mu\nu}. \label{Tmunu}
\end{equation}

In order to identify the Einstein tensor calculated with the metric connection (Christoffel symbols) and to define an effective energy-momentum tensor gathering the scalar field terms, we can use the definition of Ricci tensor  
\begin{equation}
    R_{\mu\nu} = \partial_{\nu}\Gamma^{\tau}_{\mu\tau} - \partial_{\tau}\Gamma^{\tau}_{\mu\nu} + \Gamma^{\tau}_{\sigma\mu}\Gamma^{\sigma}_{\tau\nu} - \Gamma^{\tau}_{\mu\nu}\Gamma^{\sigma}_{\sigma\tau}, \label{RicciTensor}
\end{equation}
which together with the affine connection in (\ref{WeylConnection}) results in
\begin{equation}
R_{\mu\nu} = \tilde{R}_{\mu\nu} - \tilde{\nabla}_{\mu}\phi_{,\nu} - \frac{1}{2}\phi_{,\mu}\phi_{,\nu} + \frac{1}{2}g_{\mu\nu}\left(\phi_{,\alpha}\phi^{,\alpha} - \tilde{\nabla}_{\alpha}\phi^{,\alpha}\right) \label{SplitRicciTensor}
\end{equation}
and
\begin{equation}
    R = \tilde{R} - 3\tilde{\Box}\phi + \frac{3}{2}\phi_{,\alpha}\phi^{,\alpha}, \label{SplitScalarCurvatura}
\end{equation}
where $\tilde{R}_{\mu\nu}$, $\tilde{R}$, $\tilde{\nabla}$ and $\tilde{\Box}$ denote tensors and derivatives calculated with Christoffel symbols, as usual. With these results, we can write the Einstein tensor $G_{\mu\nu} \doteq R_{\mu\nu} - \frac{1}{2}Rg_{\mu\nu}$ as
\begin{equation}
    R_{\mu\nu} - \frac{1}{2}Rg_{\mu\nu} \equiv \tilde{R}_{\mu\nu} - \frac{1}{2}\tilde{R} g_{\mu\nu} - \tilde{\nabla}_{\mu}\phi_{,\nu} - \frac{1}{2}\phi_{,\mu}\phi_{,\nu} - g_{\mu\nu}\left(\frac{1}{4}\phi_{,\alpha}\phi^{,\alpha} - \tilde{\nabla}_{\alpha}\phi^{,\alpha}\right). \label{SplitEinsteinTensor}
\end{equation}

By inserting \eqref{SplitEinsteinTensor} in \eqref{EinsteinFieldEqs}, we get
\begin{equation}
    \tilde{R}_{\mu\nu} - \frac{1}{2}\tilde{R} g_{\mu\nu} = - \mathcal{T}_{\mu\nu},
\end{equation}
where we defined
\begin{equation}
    \mathcal{T}_{\mu\nu} \doteq T_{\mu\nu}^{(\phi)} - \tilde{\nabla}_{\mu}\phi_{,\nu} - \frac{1}{2}\phi_{,\mu}\phi_{,\nu} - g_{\mu\nu}\left(\frac{1}{4}\phi_{,\alpha}\phi^{,\alpha} - \tilde{\nabla}_{\alpha}\phi^{,\alpha}\right),
\end{equation}
which by using \eqref{Tmunu} can be expressed simply as follows
\begin{equation}
    \mathcal{T}_{\mu\nu} \equiv \left(\omega - \frac{1}{2}\right)\phi_{,\mu}\phi_{,\nu} - \tilde{\nabla}_{\mu}\phi_{,\nu} - \frac{1}{2}g_{\mu\nu}\left[\left(\omega + \frac{1}{2}\right)\phi_{,\alpha}\phi^{,\alpha} - e^{-\phi}V - 2\tilde{\nabla}_{\alpha}\phi^{,\alpha} \right]. \label{effectiveTmunu}
\end{equation}
Furthermore, by performing the variation of the action with respect to the scalar field $\phi$, we obtain the following field equation,
\begin{equation}
    \tilde{\Box}\phi - \phi^{,\alpha}\phi_{,\alpha} + \frac{e^{-\phi}}{2\omega}\frac{d V}{d \phi} = 0 .\label{phiVariation}
\end{equation}    
\section{Pointlike Lagrangian and FRW and Klein-Gordon equations\label{secPointlike}}
As was done in \cite{deSouza:2013uu}, for the analysis of the cosmological aspects of the model through the Noether symmetry approach, it is a
necessary step to determine the pointlike Lagrangian corresponding to the model for a FRW metric. Let us now restrict ourselves to homogeneous and isotropic cosmological models, with the Friedman-Robertson-Walker metric spatially flat given by
\begin{equation}
ds^{2} = dt^{2}-a^{2}(t)\left[  dx^{2} +  dy^{2} +  dz^{2}  \right]  .\label{dsFW}
\end{equation}
Besides that, in order of do not spoil the homogeneity and isotropicity, we need to require that $\phi = \phi(t)$. Therefore, according to the metric (\ref{dsFW})  the kinetic term of the scalar field will reduce to
\begin{equation}
\omega g^{\mu\nu} \phi_{,\mu}\phi_{,\nu} = \omega\dot{\phi}^2, \label{CinTermFW}
\end{equation}
where dot means derivative with respect to coordinate $t$. In terms of the metric (\ref{dsFW}) the Ricci scalar will be express by
\begin{equation}
R = 6\left( \frac{\ddot{a}}{a}+ \frac{\dot{a}^{2}}{a^2}\right) + \frac{3}{2} \left( \dot{\phi}^{2} - 6\dot{\phi}\frac{\dot{a}}{a} - 2\ddot{\phi}\right).\label{RicciScalarFW}
\end{equation}
In this way, by using eqs.(\ref{RicciScalarFW}, \ref{CinTermFW}) in the eq.(\ref{Sjf}), and that $d^{4}x \sqrt{-g} =a^{3}dt d^{3}x  $, we will have
\begin{equation}
\mathcal{S}_{JF} = \mathcal{V}_{o} \int dt e^{-\phi}\left[ 6\left( a^2\ddot{a}+a\dot{a}^2 \right) +\frac{3}{2}\left(a^3\dot{\phi}^2 -6a^2\dot{a}\dot{\phi}-2a^3\ddot{\phi} \right) + \omega a^3\dot{\phi}^2 - a^{3}e^{-\phi}V\left(\phi\right) \right]\label{ARed1FW} ,
\end{equation}
where we have defined $\mathcal{V}_{o}$ as the 3-volume
\begin{equation}
\mathcal{V}_{o} \doteq \int\limits_{\mathcal{M}} d^3 x \label{Vo}.
\end{equation}
The eq.(\ref{ARed1FW}) can be integrated by parts to separate terms of total derivative. Thus, by running this procedure we will get the following reduced action
\begin{equation}
\mathcal{S}_{JF} = V_{o} \int dt\left\lbrace e^{-\phi}\left[ \left( \omega - \frac{3}{2}\right)a^3\dot{\phi}^2 + 6\left(a^2\dot{a}\dot{\phi} -a\dot{a}^2 \right)- a^{3}e^{-\phi}V\left(\phi\right) \right] + \frac{d}{dt}\left[e^{-\phi}a^2\left( 6\dot{a} - 3a\dot{\phi}\right) \right] \right\rbrace \label{ARed2FW}.
\end{equation}
The last term in the eq.(\ref{ARed2FW}) is a surface term, so that we can write
\begin{equation}
\mathcal{S}_{JF} = V_{o} \int dt\left\lbrace e^{-\phi}\left[ \left( \omega - \frac{3}{2}\right)a^3\dot{\phi}^2 + 6\left(a^2\dot{a}\dot{\phi} -a\dot{a}^2 \right) - a^{3}e^{-\phi}V\left(\phi\right)\right] \right\rbrace + \textrm{Surface term}.
\end{equation}
Thus, after neglecting surface terms, we obtain the pointlike Lagrangian
\begin{equation}
\mathcal{L}_{JF} \doteq e^{-\phi}\left[ \left( \omega - \frac{3}{2}\right)a^3\dot{\phi}^2 + 6\left(a^2\dot{a}\dot{\phi} -a\dot{a}^2 \right) - a^{3}e^{-\phi}V\left(\phi\right)\right]. \label{LagranFW}
\end{equation}

From the Euler-Lagrange equation for $a$ applied to (\ref{LagranFW}), 
\begin{equation}
\frac{d}{dt}\left( \frac{\partial \mathcal{L}}{\partial \dot{a}}\right) - \frac{\partial \mathcal{L}}{\partial a} = 0 \label{EL-equation} 
\end{equation}
we obtain the acceleration equation
\begin{equation}
\frac{\ddot{a}}{a} = -\frac{1}{6}\left(\rho_{\phi}+3 p_{\phi}\right) \label{AccelEq}
\end{equation}
where $H \doteq \frac{\dot{a}}{a}$ is the Hubble parameter. By imposing that the energy function associated with (\ref{LagranFW}) vanishes, 
\begin{equation}
E_{\mathcal{L}} \equiv \frac{\partial \mathcal{L}}{\partial \dot{a}}\dot{a} +\frac{\partial \mathcal{L}}{\partial \dot{\phi}}\dot{\phi} - \mathcal{L} = 0
\end{equation}
we have the Friedmann equation,
\begin{equation}
 H^{2} = \frac{1}{3}\rho_{\phi} \label{FriedmanEq}
\end{equation}
In equations (\ref{AccelEq}) and (\ref{FriedmanEq}) we defined the effective energy density and pressure of the scalar field as follows
\begin{eqnarray}
\rho_{\phi} & = & \frac{1}{2}\left(\omega - \frac{3}{2}\right)\dot{\phi}^{2}+\frac{1}{2}e^{-\phi}V + 3H\dot{\phi} \label{density}\\
p_{\phi} & = & \frac{1}{2}\left(\omega + \frac{1}{2}\right)\dot{\phi}^{2} - \frac{1}{2}e^{-\phi}V - 2H\dot{\phi} - \ddot{\phi} \label{pressure} 
\end{eqnarray}
in accordance to the energy-momentum tensor in \eqref{effectiveTmunu}\footnote{It is useful remember that $\rho \equiv \mathcal{T}_{00}$ and $p \equiv -\frac{1}{3}(\mathcal{T} - \mathcal{T}_{00})$ when we use a comoving frame $U^{\mu}\delta_{\mu}^{0}$ and, futhermore, we can remember the identity $\tilde{\nabla}_{\alpha}V^{\alpha} \equiv \frac{1}{\sqrt{-g}}\partial_{\alpha}(\sqrt{-g}V^{\alpha})$}.
Now, from the Euler-Lagrange equation for $\phi$ applied to (\ref{LagranFW}),
\begin{equation}
\frac{d}{dt}\left( \frac{\partial \mathcal{L}}{\partial \dot{\phi}}\right) - \frac{\partial \mathcal{L}}{\partial \phi} = 0,
\end{equation}
we get the Klein-Gordon equation as follows
\begin{equation}
    \ddot{\phi} + 3H\dot{\phi} - \dot{\phi}^2 + \frac{e^{-\phi}}{2\omega}\frac{d V}{d\phi} = 0, \label{Klein-Gordon}
\end{equation}
in accordance to the scalar field equation in \eqref{phiVariation}.
\section{Noether symmetry\label{secNoether}}
Now we are ready to employ the Noether symmetry approach to constrain the self-interaction potential. Let us consider the following infinitesimal generator of symmetry,
\begin{equation}
	\bm{X} = \alpha \frac{\partial }{\partial a} + \beta \frac{\partial }{\partial \phi} + \left( \dot{a} \frac{\partial \alpha}{\partial a} + \dot{\phi} \frac{\partial \alpha }{\partial \phi}\right)\frac{\partial}{\partial\dot{a}} + \left( \dot{a} \frac{\partial \beta}{\partial a} + \dot{\phi} \frac{\partial \beta }{\partial \phi}\right)\frac{\partial}{\partial\dot{\phi}} \label{vecfield}
\end{equation}
where $\alpha$ and $\beta$ are functions only of $a$ and $\phi$. There will exist a Noether symmetry for the point-like Lagrangian of our model if the condition
\begin{equation}
L_{\bm{X}}\mathcal{L} \equiv \bm{X}\mathcal{L} = 0 \label{symcondition}
\end{equation}
holds, i.e., if the Lie derivative of the Lagrangian with respect to the vector field $\bm{X}$ vanishes \cite{deSouza:2013uu}. By applying the symmetry condition (\ref{symcondition}) to (\ref{LagranFW}), with respect to the vector field (\ref{vecfield}), we obtain a system of coupled partial differential equations as shown bellow
\begin{eqnarray}
\left( 3\alpha -2a\beta \right) V\left( \phi \right) + a\beta V'\left( \phi \right) & = & 0 , \label{eq1}\\
\alpha + 2a\frac{\partial \alpha}{\partial a} - a \left( \beta + a \frac{\partial \beta}{\partial a}\right) & = & 0 , \label{eq2}\\
a\beta - \left( 2\alpha +a\frac{\partial \alpha}{\partial a} + a\frac{\partial \beta}{\partial a}\right) + 2\frac{\partial \alpha}{\partial \phi} - \frac{1}{3}\left(\omega -\frac{3}{2}\right)a^{2}\frac{\partial \beta}{\partial a} & = & 0 , \label{eq3}\\
\left(\omega -\frac{3}{2}\right)\left( 2a\frac{\partial\beta}{\partial \phi}-a\beta +3\alpha\right) + 6\frac{\partial \alpha}{\partial \phi} & = & 0 . \label{eq4}
\end{eqnarray}
Equation \eqref{eq1} can be written as 
\begin{equation}
    \left( \frac{3 \alpha}{2 a \beta} - 1\right) = f(\phi) 
\end{equation}
where we have defined $f(\phi)\doteq -\frac{V'(\phi)}{2V(\phi)}$. The differentiation of the left-hand side of the above equation with respect to $a$ leads to the following differential equation
\begin{equation}
    \frac{1}{\alpha}\frac{\partial \alpha}{\partial a} - \frac{1}{\beta}\frac{\partial\beta}{\partial a} = \frac{1}{a},
\end{equation}
whose solution is 
\begin{equation}
    \alpha = a \beta g(\phi) \label{sol1}
\end{equation}
where $g(\phi)$ is arbitrary function. From \eqref{eq2} together with  we \eqref{sol1} we have
\begin{equation}
    g(\phi) = \frac{\beta + a \frac{\partial \alpha }{\partial a}}{3\beta + 2a \frac{\partial \alpha }{\partial a}}. \label{funcG}
\end{equation}
The differentiation of the above equation with respect to $a$ leads to the following differential equation
\begin{equation}
    \beta \left( a \frac{\partial^2 \beta}{\partial a^2} + \frac{\partial \beta}{\partial a}\right) - a\left(\frac{\partial \beta}{\partial a} \right)^{2} = 0,
\end{equation}
whose solution is
\begin{equation}
    \beta = h(\phi) a^{n}, \label{solBeta}
\end{equation}
where $h(\phi)$ is an arbitrary function and $n$ is not specified, it can assume some value which will depend on the two remaining equations. Inserting \eqref{solBeta} into \eqref{funcG} we have that
\begin{equation}
    g(\phi) = \frac{1 + n}{3 +2n} \label{funcG2}
\end{equation}
is a constant. The equations \eqref{eq3} and \eqref{eq4} admit the solution
\begin{equation}
    \frac{d h}{d \phi} = 0,
\end{equation}
only if $n = 0$, this implies that
\begin{equation}
    \alpha = \frac{1}{3}a \beta_{0}, \label{solAlpha}
\end{equation}
which $\beta_{0}$ is a constant, and \eqref{eq1} reduces to 
\begin{equation}
    \frac{V'(\phi)}{V(\phi)} = 1 
\end{equation}
whose the solution is 
\begin{equation}
    V(\phi) = 2\Lambda e^{\phi} \label{solV}
\end{equation}
where $\Lambda$ is a constant. Because that, the action \eqref{Sjf} could take the following form
\begin{equation}
\mathcal{S}_{JF} = \int d^{4}x \sqrt{-g}e^{-\phi}\left( R + \omega \phi^{,\alpha}\phi_{,\alpha} - 2\Lambda\right).
\end{equation}
while in the Einstein frame the action \eqref{Sef} becomes
\begin{equation}
    \mathcal{S}_{EF} = \int d^{4}x \sqrt{-\bar{g}}\left( \bar{R} + \omega \phi^{,\alpha}\phi_{,\alpha} - 2\Lambda e^{\phi}\right)
\end{equation}
By the way, the conserved quantity associated with the Noether symmetry corresponding to this solution is given by \cite{deSouza:2008nj}
\begin{equation}
    \Sigma_{0} \doteq \alpha\frac{\partial \mathcal{L}}{\partial \dot{a}} + \beta\frac{\partial \mathcal{L}}{\partial \dot{\phi}}. \label{ConservQuantity}
\end{equation}
Therefore, by using \eqref{LagranFW} and \eqref{solAlpha} in \eqref{ConservQuantity}, we obtain
\begin{equation}
    \Sigma_{0} = 2\beta_{0}a^{2}e^{-\phi}\left[ \dot{a} + \left( \omega - \frac{1}{2}\right)a\dot{\phi}\right]. \label{Sigma}
\end{equation}
In the next section, we shall look for analytical solutions. 
\section{Solutions of the field equations\label{secSolution}}
To find the solutions of the field equations we need to
rewrite the pointlike Lagrangian \eqref{LagranFW} in another variables which makes integration easier. Thus, by
knowing that there is a Noether symmetry related to $V$, there must exist a coordinate transformation in the space of configuration in which one of these coordinates is \textit{cyclic}. Such a transformation obeys the following system of differential equations:
\begin{eqnarray}
\alpha \frac{\partial u}{\partial a} & + & \beta \frac{\partial u}{\partial \phi}= 0 ,\\
\alpha \frac{\partial z}{\partial a} & + & \beta \frac{\partial z}{\partial \phi} = 1.
\end{eqnarray}
where $u = u(a,\phi)$ and $z = z(a,\phi)$ are the new variables linked to the old ones, $a$ and $\phi$. In this transformation $z$ is the cyclic coordinate. It is worth to remember that due to  \eqref{solAlpha} and \eqref{solV}, we have
\begin{equation}
    \alpha = \frac{1}{3}\beta_{0}a, \;\;\;\; \beta = \beta_{0} \;\;\;\; \textrm{and} \;\;\;\; V = 2\Lambda e^{\phi} \nonumber.
\end{equation}
Thus, the system of differential equations above takes the form
\begin{eqnarray}
\frac{a}{3} \frac{\partial u}{\partial a} & + & \frac{\partial u}{\partial \phi} = 0 ,\\
\frac{a}{3} \frac{\partial z}{\partial a} & + &  \frac{\partial z}{\partial \phi} = \frac{1}{\beta_{0}},
\end{eqnarray}
whose solutions are given below
\begin{equation}
u = a^{3} e^{-\phi} \;\;\;\; \textrm{and} \;\;\;\;\
z = \frac{3}{\beta_{0}}\ln(a) .
\end{equation}
It is also useful to have expressions for $\dot{a}$ and $\dot{\phi}$,
\begin{equation}
    a \equiv e^{\beta_{0} z /3}, \;\;\;\; \dot{a} \equiv \frac{1}{3}\beta_{0}\dot{z}e^{\beta_{0}z/3}, \;\;\;\; H \equiv \frac{1}{3}\beta_ {0} \dot{z} \;\;\;\; \textrm{and} \;\;\;\;\dot{\phi} \equiv \beta_{0}\dot{z} - \frac{\dot{u}}{u}. \nonumber
\end{equation}
By taking into account these transformations, we get the following expression to \eqref{LagranFW}
\begin{equation}
    \mathcal{L} = k_{1}\dot{u}\dot{z} + k_{2}u\dot{z}^{2}+ k_{3}\frac{\dot{u}^2}{u} - 2 u \Lambda .\label{LagranTransf}
\end{equation}
where we have defined the parameters 
\begin{equation}
   k_{1} \doteq - 2\beta_{0}\left(\omega - \frac{1}{2}\right), \;\;\;\; k_{2} \doteq \beta_{0}^{2}\left(\omega - \frac{1}{6}\right) \;\;\;\; \textrm{and} \;\;\;\;k_{3} \doteq \left(\omega - \frac{3}{2}\right). \label{defks}
\end{equation}
Now, from the Euler-Lagrange equations associated with the Lagrangian \eqref{LagranTransf}, we obtain the field equations in the new variables, namely,
\begin{eqnarray}
k_{1}\dot{u} + 2k_{2}u\dot{z} = \Sigma_{0} \label{NewEq1}\\
2k_{3} \frac{\ddot{u}}{u} + k_{1}\ddot{z} - k_{2}\dot{z}^{2} - k_{3}\frac{\dot{u}^{2}}{u^{2}} + 2\Lambda = 0 \label{NewEq2}
\end{eqnarray}
where $\Sigma_{0}$ is the constant of motion \eqref{Sigma} rewritten in the new variables. This agrees with the fact that $z$ been a cyclic coordinate, implying that the momentum canonically conjugate to the $z$ is conserved
\begin{equation}
    p_{z} \doteq \frac{\partial \mathcal{L}}{\partial \dot{z}} \implies p_{z} = k_{1}\dot{u} + 2k_{2}u\dot{z} \equiv \Sigma_{0}.
\end{equation}
Therefore,
\begin{eqnarray}
\frac{d p_{z}}{dt} &=& 0,\\
\frac{d p_{u}}{dt} &=& k_{2}\dot{z}^2 - k_{3}\frac{\dot{u}^{2}}{u^{2}} - 2\Lambda.
\end{eqnarray}
The energy function associated with the Lagrangian \eqref{LagranTransf} provides another equation \eqref{FriedmanEq}
\begin{equation}
    k_{3}\frac{\dot{u}^{2}}{u^{2}} + k_{1}\frac{\dot{u}}{u}\dot{z} + k_{2}\dot{z}^{2} + 2\Lambda= 0 ,\label{NewFriedEq}
\end{equation}
which is equivalent to the Friedmann equation in the old variables. 

The equations \eqref{NewEq1}, \eqref{NewEq2} and \eqref{NewFriedEq} comprehend a system of three differential equations for two dynamical variables, $u$ and $z$. To obtain the solutions of these equations, we shall isolate $\dot{z}$ from \eqref{NewEq1}, 
\begin{equation}
    \dot{z} = \frac{\Sigma_{0} - k_{1}\dot{u}}{2k_{2}u} \label{dotz}
\end{equation}
and to substitute $\dot{z}$ in \eqref{NewEq2}, 
\begin{equation}
    \left(k_{3}-\frac{k_{1}^{2}}{4k_{2}}\right)\dot{u}^{2} + 2\Lambda u^{2} + \frac{\Sigma_{0}^2}{4k_{2}} = 0.
\end{equation}
We can write the equation above in the following canonical form
\begin{equation}
    \dot{u}^2 + \lambda u^2 + \sigma = 0 \label{ODE1}
\end{equation}
where we have defined the parameters $\lambda$ and $\sigma$ as
\begin{eqnarray}
\lambda &\doteq& \frac{8\Lambda k_{2}}{\left(4k_{2}k_{3}-k_{1}^{2}\right)} \equiv \frac{\Lambda}{2\omega}(1-6\omega)\\
\nonumber \\
\sigma &\doteq& \frac{\Sigma_{0}^2}{\left(4k_{2}k_{3}-k_{1}^{2}\right)} \equiv -\frac{3\Sigma_{0}^2}{8\beta_{0}^2\omega},
\end{eqnarray}
which we used the definitions in \eqref{defks}. The ODE \eqref{ODE1} allows the solution given below
\begin{equation}
    u(t) = \pm \sqrt{-\frac{\sigma}{\lambda}}\sin\left[\sqrt{\lambda}\left(t\pm C_{1}\right)\right], \label{GenSolU}
\end{equation}
where $C_{1}$ is arbitrary constant of integration. We shall take $C_{1} = 0$. In order to get $z(t)$ by integrating \eqref{dotz}, we will handle the cases for each $\lambda \neq 0$ sign.
\subsection{Case \texorpdfstring{$\lambda<0$}{}:}
This case can be considered if
\begin{equation*}
    \lambda < 0 \implies \begin{cases}
                \Lambda > 0 &\text{and } \omega \in \lbrace \left(-\infty, 0\right) \cup \left(1/6, \infty \right)\rbrace\\
                            &\text{ or }\\
                \Lambda < 0 &\text{and } \omega \in \left(0, 1/6\right)
            \end{cases}
\end{equation*}
Therefore, in this case, \eqref{GenSolU} becomes,
\begin{equation}
    u(t) = \pm \frac{\Sigma_{0}}{2\beta_{0}}\sqrt{\frac{3}{|2\omega\lambda|}}\sinh\left( \sqrt{|\lambda|}t\right). \label{SolULambdaNeg}
\end{equation}
By considering the equation above in \eqref{dotz}, we may obtain
\begin{equation}
    \dot{z} = \frac{\Lambda}{\omega\lambda\beta_{0}}\left[ \frac{(3-6\omega)}{2}\sqrt{|\lambda|}\coth\left(\sqrt{|\lambda|}t\right) \mp \sqrt{6|\omega\lambda|} \textrm{csch}\left(\sqrt{|\lambda|}t\right)
    \right], \label{dotzLambdaNeg}
\end{equation}
which can be integrated in $t$, 
\begin{equation}
 z(t) = \frac{\Lambda}{\omega\lambda\beta_{0}}\left\lbrace \frac{(3-6\omega)}{2}\ln\left[\sinh\left(\sqrt{|\lambda|}t\right)\right] \pm \sqrt{6|\omega|}\ln\left[ \coth\left(\frac{\sqrt{|\lambda|}}{2}t\right)\right]\right\rbrace.
 \end{equation}
In terms of the original variables $a(t)$ and $\phi(t)$, we have
\begin{equation}
    a(t) = \exp\left\lbrace\frac{\Lambda}{3\omega\lambda}\left[ \frac{(3-6\omega)}{2}\ln\left[\sinh\left(\sqrt{|\lambda|}t\right)\right] \pm \sqrt{6|\omega|}\ln\left[ \coth\left(\frac{\sqrt{|\lambda|}}{2}t\right)\right]\right]\right\rbrace \label{sola1}
\end{equation}
and, 
\begin{equation}
    \phi(t) = \frac{\Lambda}{\omega\lambda}\left\lbrace \ln\left[\sinh\left(\sqrt{|\lambda|}t\right)\right] \pm \sqrt{6|\omega|}\ln\left[ \coth\left(\frac{\sqrt{|\lambda|}}{2}t\right)\right]\right\rbrace.
\end{equation}

\subsection{Case \texorpdfstring{$\lambda > 0$}{}:}
This case can be considered if
\begin{equation*}
 \lambda > 0 \implies \begin{cases}
                \Lambda > 0 &\text{and } \omega \in \left(0, 1/6\right)\\
                            &\text{ or }\\
                \Lambda < 0 &\text{and } \omega \in \lbrace \left(-\infty, 0\right) \cup \left(1/6, \infty \right)\rbrace
            \end{cases}
\end{equation*}
Therefore, in this case, \eqref{GenSolU} becomes,
\begin{equation}
    u(t) = \pm\frac{\Sigma_{0}}{2\beta_{0}}\sqrt{\frac{3}{2\omega\lambda}}\sin\left( \sqrt{\lambda}t\right). \label{SolULambdaPos}
\end{equation}
By considering the equation above in \eqref{dotz}, we may obtain
\begin{equation}
    \dot{z} = \frac{\Lambda}{\omega\lambda\beta_{0}}\left[ \frac{(3-6\omega)}{2}\sqrt{\lambda}\cot\left(\sqrt{|\lambda|}t\right) \mp \sqrt{6|\omega\lambda|} \csc\left(\sqrt{|\lambda|}t\right),
    \right] \label{dotzLambdaPos}
\end{equation}
which can be integrated in $t$, 
\begin{equation}
 z(t) = \frac{\Lambda}{\omega\lambda\beta_{0}}\left\lbrace \frac{(3-6\omega)}{2}\ln\left[\sin\left(\sqrt{|\lambda|}t\right)\right] \pm \sqrt{6|\omega|}\ln\left[ \cot\left(\frac{\sqrt{|\lambda|}}{2}t\right)\right]\right\rbrace.
 \end{equation}
In terms of the original variables $a(t)$ and $\phi(t)$, we have
\begin{equation}
    a(t) = \exp\left\lbrace\frac{\Lambda}{3\omega\lambda}\left[ \frac{(3-6\omega)}{2}\ln\left[\sin\left(\sqrt{|\lambda|}t\right)\right] \pm \sqrt{6|\omega|}\ln\left[ \cot\left(\frac{\sqrt{|\lambda|}}{2}t\right)\right]\right]\right\rbrace
\end{equation}
and, 
\begin{equation}
    \phi(t) = \frac{\Lambda}{\omega\lambda}\left\lbrace \ln\left[\sin\left(\sqrt{|\lambda|}t\right)\right] \pm \sqrt{6|\omega|}\ln\left[ \cot\left(\frac{\sqrt{|\lambda|}}{2}t\right)\right]\right\rbrace.
\end{equation}
In a nutshell, we have this set of solutions,
\begin{eqnarray}
a_{\lambda < 0}^{(\pm)} & = & \exp\left\lbrace\frac{\Lambda}{3\omega\lambda}\left[ \frac{(3-6\omega)}{2}\ln\left[\sinh\left(\sqrt{|\lambda|}t\right)\right] \pm \sqrt{6|\omega|}\ln\left[ \coth\left(\frac{\sqrt{|\lambda|}}{2}t\right)\right]\right]\right\rbrace \\
\phi_{\lambda < 0}^{(\pm)} & = & \frac{\Lambda}{\omega\lambda}\left\lbrace \ln\left[\sinh\left(\sqrt{|\lambda|}t\right)\right] \pm \sqrt{6|\omega|}\ln\left[ \coth\left(\frac{\sqrt{|\lambda|}}{2}t\right)\right]\right\rbrace, \\
a_{\lambda > 0}^{(\pm)} & = & \exp\left\lbrace\frac{\Lambda}{3\omega\lambda}\left[ \frac{(3-6\omega)}{2}\ln\left[\sin\left(\sqrt{|\lambda|}t\right)\right] \pm \sqrt{6|\omega|}\ln\left[ \cot\left(\frac{\sqrt{|\lambda|}}{2}t\right)\right]\right]\right\rbrace, \\
\phi_{\lambda > 0}^{(\pm)} & = & \frac{\Lambda}{\omega\lambda}\left\lbrace \ln\left[\sin\left(\sqrt{|\lambda|}t\right)\right] \pm \sqrt{6|\omega|}\ln\left[ \cot\left(\frac{\sqrt{|\lambda|}}{2}t\right)\right]\right\rbrace.
\end{eqnarray}
Besides that, we can write $a(\phi)$ simply how,
\begin{equation}
    a(\phi) = \exp\left[-\left(\omega - \frac{1}{2}\right)\phi\right]
\end{equation}
\section{Cosmological solutions in each frame\label{secCosmologicalSolutions}}
In this section we shall give the expressions for the scale factor, Hubble and deceleration parameters in the Jordan and Einstein frames. The solutions for $\lambda>0$ in the Jordan frame give oscillatory behaviors which are not interested in the cosmological sense so that we shall neglect them in this section.  
\subsection{Jordan frame}
On the basis of  the solutions for the scale factor in the Jordan frame below
\begin{equation}\label{j1}
    a_{\lambda < 0}^{(\pm)} = \exp\left\lbrace\frac{\Lambda}{3\omega\lambda}\left[ \frac{(3-6\omega)}{2}\ln\left[\sinh\left(\sqrt{|\lambda|}t\right)\right] \pm \sqrt{6|\omega|}\ln\left[ \coth\left(\frac{\sqrt{|\lambda|}}{2}t\right)\right]\right]\right\rbrace,
\end{equation}
we can obtain the following Hubble parameter 
\begin{equation}\label{j2}
    H_{\lambda < 0 }^{(\mp)} = \frac{\Lambda\sqrt{\lambda}}{3\omega\lambda}\left[ \frac{(3-6\omega)}{2}\coth\left(\sqrt{|\lambda|}t\right) \mp \sqrt{6|\omega|}\textrm{csch}\left(\sqrt{|\lambda|}t\right)\right].
\end{equation}
Furthermore by considering the definition of the deceleration parameter
\begin{equation}
    q \doteq - \frac{a\ddot a }{\dot{a}^2} \equiv - \left(\frac{\dot H}{H^2} + 1\right),
\end{equation}
its expression becomes
\begin{equation}\label{j3}
    q_{\lambda < 0 }^{(\mp)} = \frac{3\omega\lambda}{\Lambda}
    \frac{\left[\frac{(3-6\omega)}{2} \mp \sqrt{6\omega}\cosh\left(\sqrt{|\lambda|}t\right) \right]\textrm{csch}^2\left(\sqrt{|\lambda|}t\right)}
    {\left[\frac{(3-6\omega)}{2}\coth\left(\sqrt{|\lambda|}t\right) \mp \sqrt{6|\omega|}\textrm{csch}\left(\sqrt{|\lambda|}t\right)\right]^2} - 1.
\end{equation}
\subsection{Einstein frame}
In the Einstein frame  the scale factor is obtained through the Weyl transformation
\begin{equation}
    \bar{a} = ae^{-\phi/2},
\end{equation}
which implies 
\begin{equation}\label{e1}
    \bar{a}_{\lambda < 0}^{(\mp)} = \exp \left\lbrace -\frac{2\omega}{(1-6\omega)} \ln\left[\sinh\left(\sqrt{|\lambda|}t\right)\right] \mp \frac{2\omega}{(1-6\omega)\sqrt{6\omega}}\ln\left[\coth\left(\frac{\sqrt{|\lambda|}}{2}t\right)\right] \right\rbrace.
\end{equation}
From the knowledge of the scale factor the Hubble and deceleration parameters can be obtained, yielding
\begin{equation}\label{e2}
    \bar{H}_{\lambda < 0}^{(\pm)} = \frac{\textrm{csch}\left(\sqrt{|\lambda|}t\right)}{(1-6\omega)}\left[ \pm \sqrt{\frac{2}{3}|\lambda\omega|} - 2\omega\sqrt{|\lambda|}\cosh\left(\sqrt{|\lambda|}t\right)\right] ,
\end{equation}
\begin{equation}\label{e3} 
   \bar{q}_{\lambda < 0}^{(\mp)} = \frac{(1-6\omega)}{2\omega}\frac{\left[\pm\sqrt{6\omega}\cosh\left(\sqrt{|\lambda|}t\right)-6\omega\right]}{\left[1\mp \sqrt{6\omega}\cosh\left(\sqrt{|\lambda|}t\right)\right]^2} -1
\end{equation}

Here it is important to remember that the correspondence between  the scale factors in both frames are
\begin{equation}
    a_{\textrm{Jordan}}^{(+)} \Longleftrightarrow \bar{a}_{\textrm{Einstein}}^{(-)},\qquad
     a_{\textrm{Jordan}}^{(-)} \Longleftrightarrow \bar{a}_{\textrm{Einstein}}^{(+)}.
\end{equation}

\section{A particular case: \texorpdfstring{$\omega = 1/2$}{} \label{secParticularCase}}
We would like to analyze the solutions in the case where $\omega = 1/2$, since it is a very common case in the literature, a minimally coupled scalar-tensor theory in the Einstein frame. 

To begin with, let us write below the solutions for the scale factor (\ref{j1}), Hubble parameter (\ref{j2}) and deceleration parameter (\ref{j3}) in the Jordan frame, with $\omega = 1/2$:
\begin{eqnarray}
a_{\lambda < 0}^{(\mp)} = \left[ \coth\left(\sqrt{\frac{\Lambda}{2}}t\right)\right]^{\mp\frac{1}{\sqrt{3}}}\\
H_{\lambda < 0 }^{(\pm)} = \pm\sqrt{\frac{2\Lambda}{3}}\textrm{csch}\left(\sqrt{2\Lambda}t\right)\\
q_{\lambda < 0 }^{(\pm)} = \pm \sqrt{3}\cosh\left(\sqrt{2\Lambda}t\right) - 1.
\end{eqnarray}
In the Einstein frame the corresponding set of expressions which follows from (\ref{e1}), (\ref{e2}) and (\ref{e3}) for $\omega=1/2$ reads
\begin{eqnarray}
    \bar{a}_{\lambda < 0}^{(\pm)} = \left[\sinh\left(\sqrt{2\Lambda}t\right)\right]^{\frac{1}{2}}\cdot \left[\coth\left(\sqrt{\frac{\Lambda}{2}}t\right)\right]^{\pm \frac{1}{2\sqrt{3}}},\\
    \bar{H}_{\lambda < 0}^{(\mp)} = \textrm{csch}\left(\sqrt{2\Lambda}t\right)\left[ \sqrt{\frac{\Lambda}{2}}\cosh\left(\sqrt{2\Lambda}t\right) \mp \sqrt{\frac{\Lambda}{6}} \right],\\
    \bar{q}_{\lambda < 0}^{(\mp)} = \frac{\left[6 \mp 2\sqrt{3}\cosh\left(\sqrt{2\Lambda}t\right)\right]}{\left[1 \mp\sqrt{3}\cosh\left(\sqrt{2\Lambda}t\right)\right]^{2}} - 1.
\end{eqnarray}
\subsection{Jordan Frame vs. Einstein Frame}
Here we shall analyse the solutions which are compatible with an expanding universe from the expressions given above for the case of $\omega=1/2$. For the Jordan frame only the solution for the scale factor  $a^{(-)}_{\lambda<0}$ implies an expanding universe. On the other hand, in the Einstein frame  both solutions for the scale factor are possible solutions  for an expanding universe.  We shall analyse here the scale factor $\bar{a}^{(+)}_{\lambda<0}$ in the Einstein frame since it corresponds to the scale factor $a^{(-)}_{\lambda<0}$ in the Jordan frame.
\begin{figure}[H]
\centering
\includegraphics[scale = 0.6]{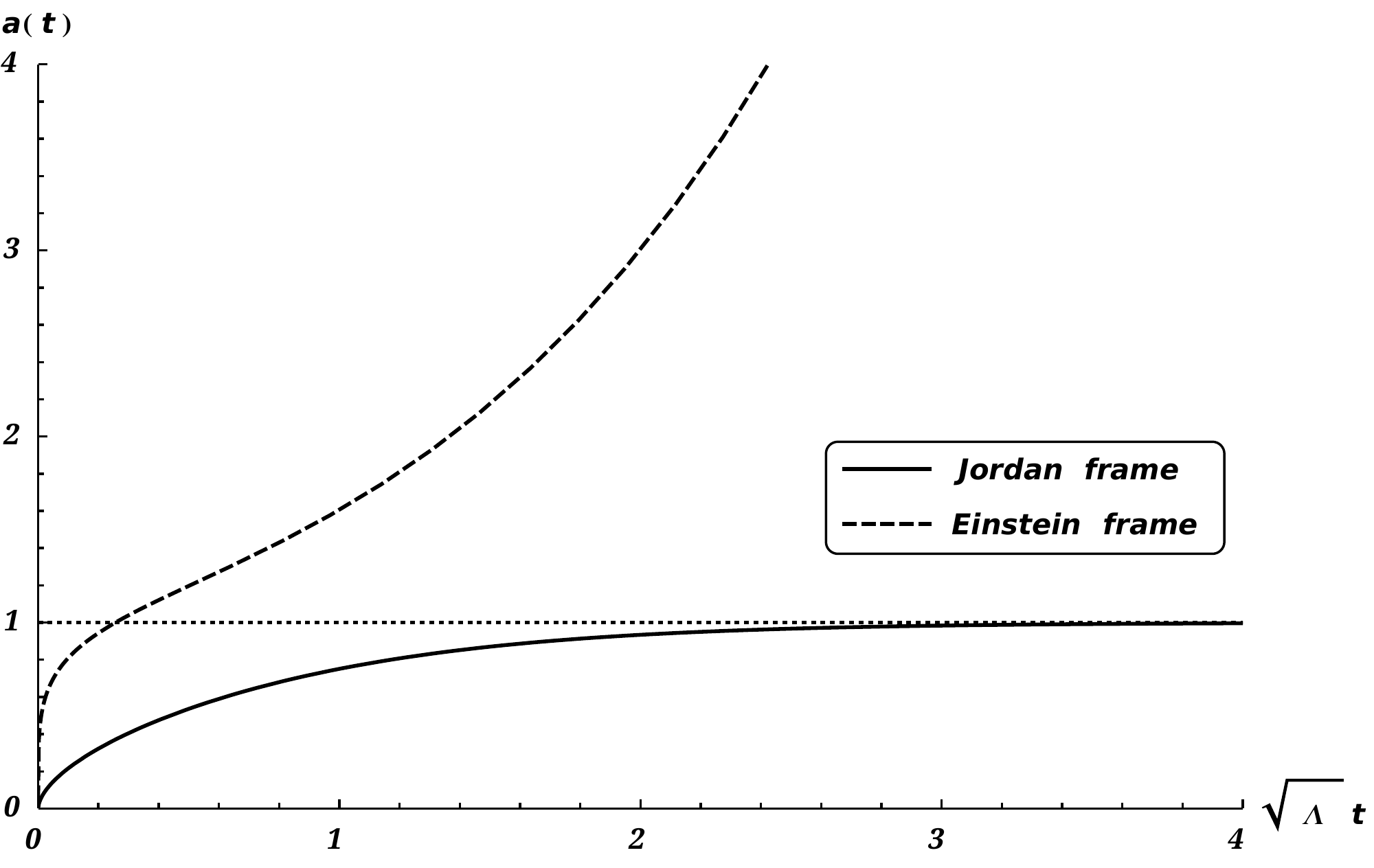}
\caption{Scale factors $a^{(-)}_{\lambda<0}$ and $\bar{a}^{(+)}_{\lambda<0}$ as functions of time $\sqrt\Lambda t$.}
\end{figure} 
In Figure 1 it is plotted the scale factors in the Jordan frame $a^{(-)}_{\lambda<0}$and in the Einstein frame $\bar{a}^{(+)}_{\lambda<0}$ as functions of time $\sqrt\Lambda t$. We infer from this figure that in the Einstein frame the scale factor increases with time, while in the Jordan frame it  grows but for large time values the  scale factor tends to constant value of a stationary universe. This behavior can be understood by analysing the scale factor velocity $\dot a(t)/\sqrt\Lambda$ as function of time $\sqrt\Lambda t$ in Figure 2. We see that the scale factor velocity in the Jordan frame decreases with time and goes to zero at large  time values. The scalar factor velocity in the Einstein frame initially decreases with time but from a certain time further it grows.  
\begin{figure}[H]
\centering
\includegraphics[scale = 0.6]{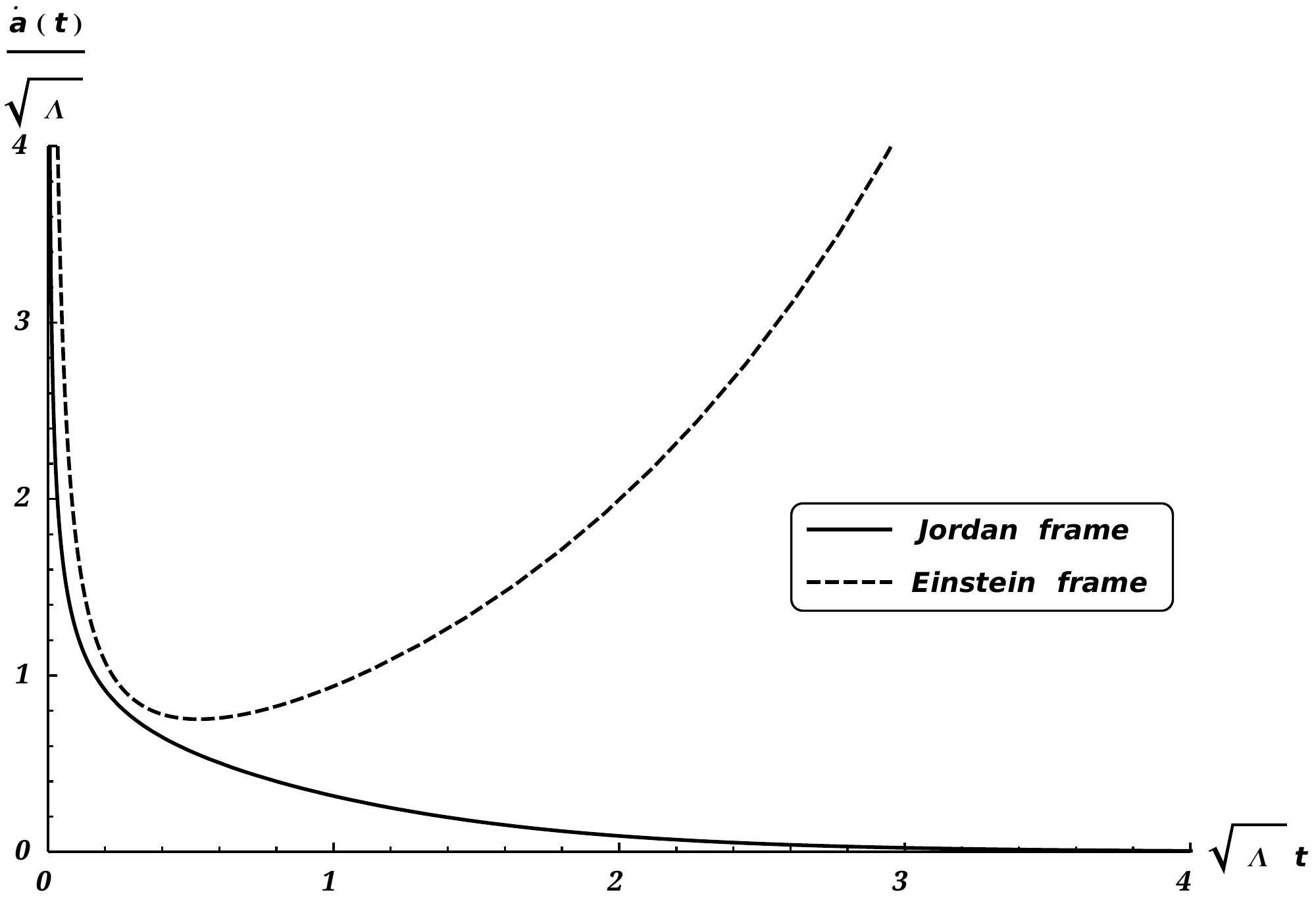}
\caption{Scale factor velocities $\dot{a}^{(-)}_{\lambda<0}/\sqrt{\Lambda}$ and $\dot{\bar{a}}^{(+)}_{\lambda<0}/\sqrt{\Lambda}$ as functions of time $\sqrt\Lambda t$.}
\end{figure} 
Figure 3 shows the behavior of the deceleration parameter $q(t)$ as function of time $\sqrt\Lambda t$ in both frames. We conclude from this figure that in the Jordan frame  the deceleration parameter has a positive  sign, which  may be interpreted as a matter dominated era. In the Einstein frame the behavior of the deceleration parameter is different from that of the Jordan frame. At the begin the deceleration parameter has positive sign and evolves to a negative sign. Here it may be interpreted to an exit of a matter dominated period to a dark energy era.
\begin{figure}[H]
\centering
\includegraphics[scale = 0.6]{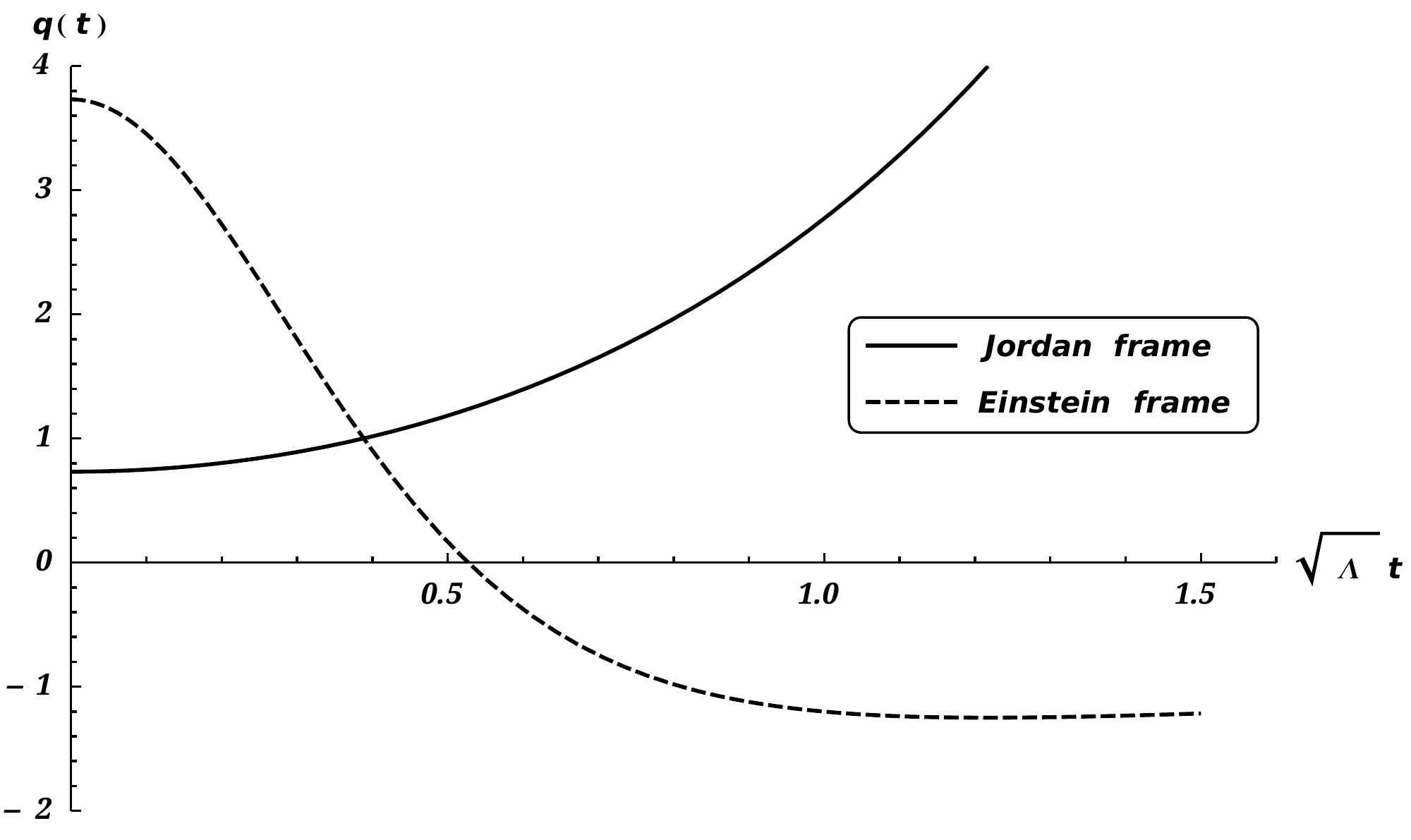}
\caption{Deceleration parameters $q^{(+)}_{\lambda<0}$ and $\bar{q}^{(-)}_{\lambda<0}$ as functions of time $\sqrt\Lambda t$.}
\end{figure}
It is clear that in this work we introduced only one constituent which is the scalar field. In order to have a better insight of the cosmological behavior one should add a matter constituent. This will be the subject of a next investigation.

\section{Conclusions\label{secConclusions}}
 In this work we analyzed a model with a scalar field minimally coupled to gravity. We started with the action in the Einstein frame and obtained the action in the Jordan frame through the use of the Weyl transformations. The field equations in the Jordan frame were obtained from the Palatini variation method. By restricting to a plane Friedman-Robertson-Walker  metric  the point-like Lagrangian and the equations of Friedmann, acceleration and Klein-Gordon were obtained. The Noether symmetry method was used to determine the self-interaction potential of the scalar field. From the solution of the field equations the scale factor, the Hubble and deceleration parameters were obtained in the Jordan frame and the corresponding ones in the Einstein frame were determined by the use of Weyl transformations. The cosmological solutions were obtained in case where the coupling constant of the scalar field $\omega=1/2$ which corresponds to a the case of a minimally coupled scalar field in the Einstein frame.  It was show that in the Jordan frame the scalar factor grows with time but tends to a constant value at large times, i. e. evolving into a stationary universe. Furthermore, its deceleration parameter has a positive sign, which may be interpreted as a  matter dominated era. In the Einstein frame the scale factor grows with time and the deceleration parameter evolves from a positive sign  to a negative one, which may be interpreted as a transition from a matter dominated period to a dark energy era.   
\vspace{6pt} 
\acknowledgments{ABB would like to thank María Laura Pucheu and Giancarlo Camilo for helpful discussions. This work was supported
by Conselho Nacional de Desenvolvimento Científico e Tecnológico (CNPq), grants Nos. 152124/2019-5 (ABB) and 304054/2019-4 (GMK).}

\bibliographystyle{utphys.bst} 
\bibliography{refs.bib}
\end{document}